%% file: main.tex
\documentclass[9pt, twocolumn]{article}

\usepackage{graphicx}
\usepackage{lipsum}  
\usepackage{xcolor}
\usepackage{booktabs}
\usepackage{multirow}
\usepackage{array}
\usepackage{url}
\usepackage{amsmath}

\usepackage[letterpaper,top=0.5in,bottom=0.9in,left=0.45in,right=0.45in,columnsep=0.22in]{geometry}

\graphicspath{ {images/} }

\title{\LARGE \bf
Lost in Transcription: \\ 
How Speech-to-Text Errors Derail Code Understanding
}

\author{
Jayant Havare$^{1}$,
Ashish Mittal$^{2}$,
Srikanth Tamilselvam$^{2}$,
Ganesh Ramakrishnan$^{1}$\\[4pt]
\small
$^{1}$IIT Bombay, Mumbai, India \quad
$^{2}$IBM Research, India\\
\texttt{\{jayantcse, ganesh\}@cse.iitb.ac.in} \quad
\texttt{\{arakeshk, srikanth.tamilselvam\}@in.ibm.com}
}

\begin{document}

\maketitle
\thispagestyle{empty}
\pagestyle{empty}

\input{abstract}

\paragraph{Keywords}
Code understanding, Speech interfaces, Automatic Speech Recognition, Multilingual programming, Large Language Models, Code-aware transcription, Indic languages, Program comprehension.

\input{introduction}

\input{motivation}

\input{related_work}

\input{framework}

\input{empirical_setup}

\input{llm_effectiveness}

\input{challenges}

\input{metrics}

\input{results}

\input{multimodal}

\input{validity}

\input{conclusion}

\medskip
 
\bibliographystyle{unsrt}  
\bibliography{reference}

\end{document}

%% file: abstract.tex
\begin{abstract}
Code understanding is a foundational capability in software engineering tools and developer workflows. However, most existing systems are designed for English-speaking users interacting via keyboards, which limits accessibility in multilingual and voice-first settings, particularly in regions like India. Voice-based interfaces offer a more inclusive modality, but spoken queries involving code present unique challenges due to the presence of non-standard English usage, domain-specific vocabulary, and custom identifiers such as variable and function names, often combined with code-mixed expressions. In this work, we develop a multilingual speech-driven framework for code understanding that accepts spoken queries in a user’s native language, transcribes them using Automatic Speech Recognition (ASR), applies code-aware ASR output refinement using Large Language Models (LLMs), and interfaces with code models to perform tasks such as code question answering and code retrieval through benchmarks such as CodeSearchNet, CoRNStack, and CodeQA. Focusing on four widely spoken Indic languages and English, we systematically characterize how transcription errors impact downstream task performance. We also identified key failure modes in ASR for code and demonstrated that LLM-guided refinement significantly improves performance across both transcription and code understanding stages. Our findings underscore the need for code-sensitive adaptations in speech interfaces and offer a practical solution for building robust, multilingual voice-driven programming tools.
\end{abstract}

%% file: introduction.tex
\section{Introduction}

The ability to understand the structure, semantics, and behavior of computer programs is a fundamental skill in programming education and practice. With the advent of artificial intelligence (AI), various chat-based assistants have emerged that help a learner interactively understand the code \cite{cursor_ai, visual_studio_code}. However, most of these tools assume English proficiency and rely on text-based, keyboard-driven interaction. This design excludes a large population of learners in multilingual regions such as India, where many students transition directly from regional language schooling to English-medium computer science curricula. For such learners, limited English fluency and typing skills create an additional barrier to mastering programming concepts. Previous work has highlighted the need for inclusive systems that support multilingual learners~\cite{reitmaier2022opportunities, nigatu2024low}, but few tools address this gap in the context of understanding code.

\begin{figure}
    \centering
    \includegraphics[width=\linewidth]{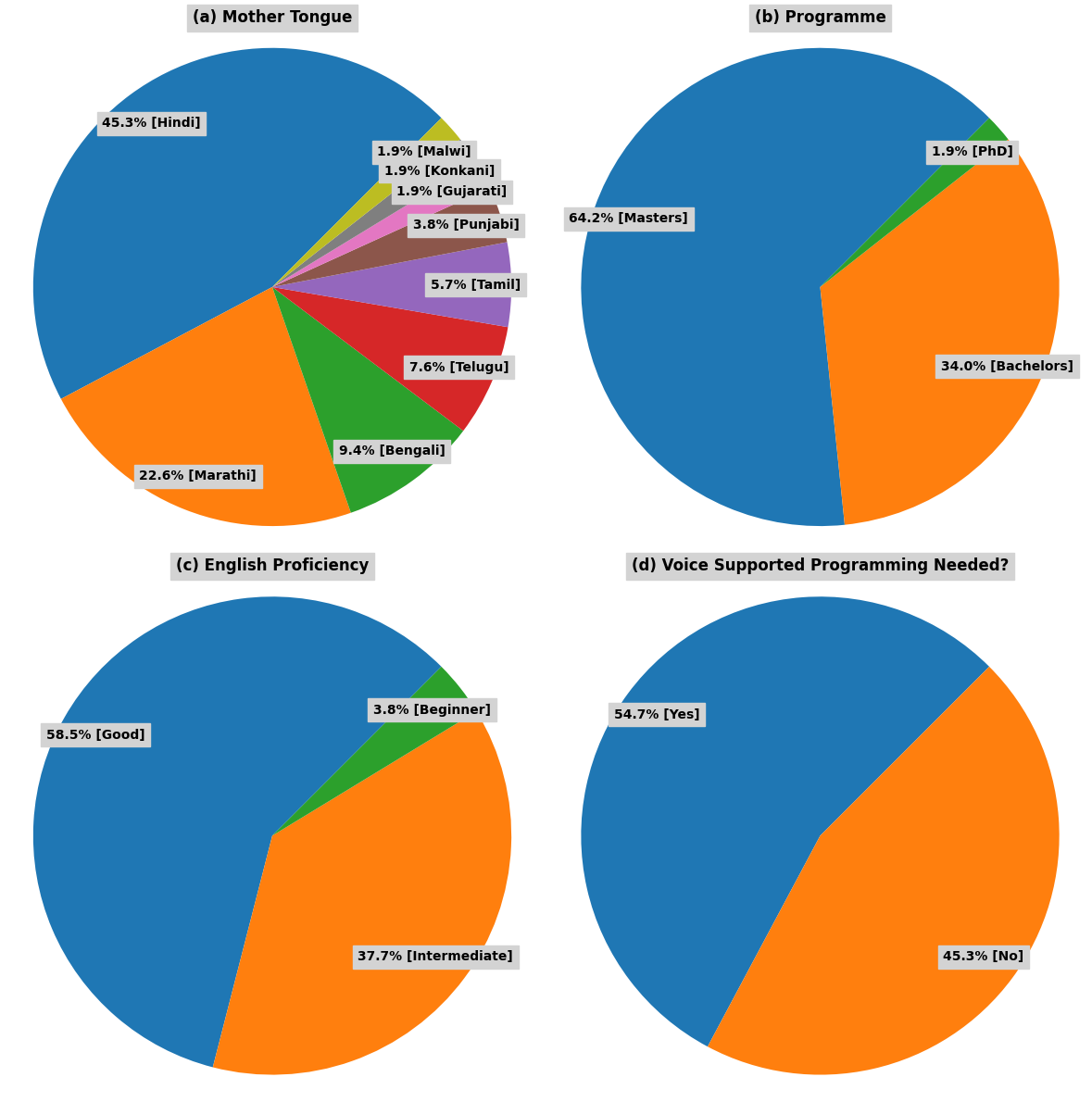}
    \caption{Survey Results: Preferences for Language and Code Understanding Modalities}
    \label{fig:survey}
\end{figure}

\begin{figure*}[t]
  \centering
  \includegraphics[width=\textwidth]{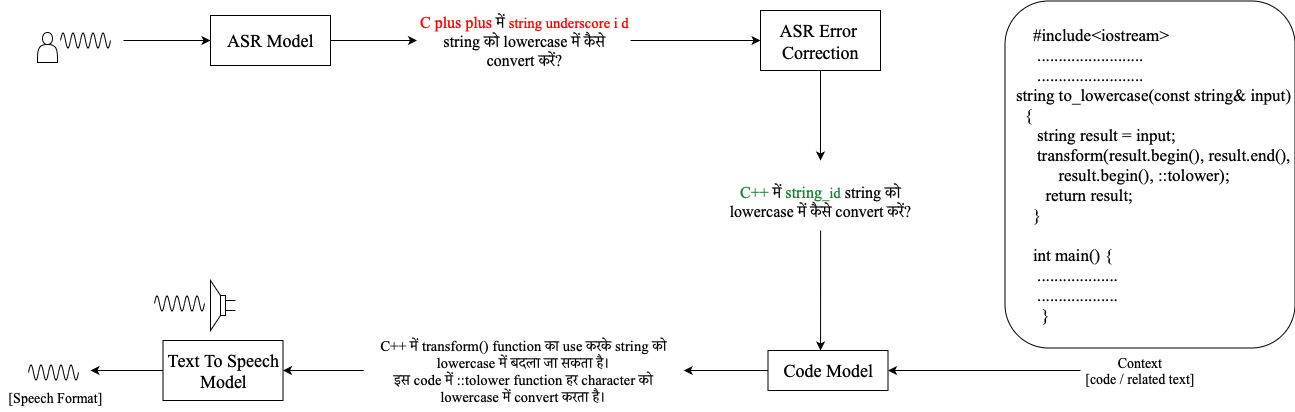}
  \caption{Multilingual ASR cascade Framework for voice-based code understanding}
  \label{fig:fullwidth}
\end{figure*}

To better understand the challenges faced by beginner programmers in multilingual contexts, we conducted a survey with 53 engineering students from multiple Indian states in order to capture learners with a wide range of mother tongues. The survey captured information about participants' academic background, self-assessed coding proficiency, comfort with English as a medium of interaction and how students perceived the utility of voice-assisted programming. As shown in Figure~\ref{fig:survey}, all participants reported their mother tongue other than English. Moreover, 41.5\% of the students did not consider themselves proficient in English; specifically, 37.7\% identified as intermediate and 3.8\% as beginners. Notably, 54.7\% of respondents indicated a preference for a voice-based interactive tool over a traditional text-based tool, while 49.1\% expressed a preference for a programming tool with multilingual support. Further details on the survey design and analysis are provided in section~\ref{motivation}.

These findings highlight a growing need for multilingual and multimodal programming support. In particular, voice-based interfaces provide a more natural and accessible way to interact with programming systems, which can help beginners learn programming and understand code more easily~\cite{paudyal2020voiceye,paudyal2022inclusive}. However, building such systems is non-trivial because spoken code-related queries are often code-mixed, syntactically irregular, and rich in out-of-vocabulary terms such as identifiers, symbols, and API names. Automatic Speech Recognition (ASR) systems, which convert spoken audio into text, frequently misinterpret such inputs~\cite{unni2022adaptive,radford2023robust,unni2023improving}, and these transcription errors propagate downstream, which eventually degrades the performance of LLMs used for code understanding~\cite{pan2024lost}. 

In this work, we present a speech-driven framework for multilingual code understanding (Figure~\ref{fig:fullwidth}). Our system accepts spoken queries in a user’s native language, transcribes them using ASR, refines the transcription via \emph{code-aware LLM-guided processing}, and finally uses a code model to generate relevant responses. The final output is delivered both in text and audio, enabling accessible feedback in the user’s preferred language.

Our study systematically investigates the robustness of our framework across four Indian languages (Hindi, Bengali, Tamil, and Gujarati) and English. We identify how transcription failures affect downstream code understanding tasks and evaluate the effectiveness of \emph{code-aware ASR output refinement} in mitigating these errors. Our contributions include:

\begin{itemize}
    \item A multilingual, voice-based system for program understanding that integrates ASR, LLM-based refinement, and code model reasoning;
    \item Failure modes taxonomy in ASR transcription for code-related speech, grounded in empirical observations;
    \item A prompt-guided refinement strategy that improves transcription fidelity and downstream performance;
    \item A thorough evaluation across five languages and two downstream tasks: code question answering and code retrieval using benchmarks like CodeSearchNet, CoRNStack, and CodeQA.
\end{itemize}

To guide our investigation, we pose the following research questions:

\begin{itemize}
    \item \textbf{RQ1: Impact on Code Understanding.} How do ASR transcription errors affect downstream code understanding tasks, such as code question answering and retrieval?
    
    \item \textbf{RQ2: Characterization of Bugs.} What are the common failure patterns introduced by ASR systems when transcribing code-related speech?
    
    \item \textbf{RQ3: Low-Resource Language Impact.} How do ASR and LLM components perform across low-resource Indic languages such as Gujarati and Tamil?
    
    \item \textbf{RQ4: LLM-Guided Refinement.} To what extent can Large Language Models refine ASR transcriptions of code-related queries and improve the performance of downstream code understanding tasks?

\end{itemize}

Our results show that even state-of-the-art ASR systems frequently misrecognize spoken code queries in multilingual settings, significantly degrading the performance of downstream code understanding tasks. However, LLM-guided refinement substantially improves both transcription quality and task accuracy.

%% file: motivation.tex
\section{Motivation}
\label{motivation}

In this section, we present the details and findings of a survey conducted with 53 first-year undergraduate students from diverse linguistic backgrounds in India. The survey was designed to assess students' language proficiency, learning preferences, and common use cases when interacting with code. 

The survey questions can be found in table \ref{tab:survey}.
These responses reveal three key trends:\\
(1) No one has English as their first language, with nearly half listing Hindi as their mother tongue. \\
(2) Only 58.5\% report being proficient in English, with others identifying as beginner or intermediate. \\
(3) More than half of the students prefer voice-based interaction for exploring or querying code.

\begin{table}[h]
\centering
\caption{Selected survey questions and responses}
\label{tab:survey}
\small
\begin{tabular}{@{}p{4.8cm}p{2.8cm}@{}}
\textbf{Survey Question} & \textbf{Top Response} \\ 
What is your mother tongue? & Hindi (45.3\%) \\
How proficient are you in English? & Good (58.5\%) \\
Do you prefer voice-based interfaces for learning code? & Yes (54.7\%) \\
What tasks do you frequently perform while learning code? & Code Q\&A, Code Retrieval \\
Would you benefit from multilingual programming support? & Yes (49.1\%) \\
\end{tabular}
\end{table}

These findings underscore the need for multilingual, speech-driven programming tools that support natural interaction and accommodate regional language preferences. In particular, code understanding tasks such as question answering over code snippets or retrieving relevant code examples were frequently cited by students, making them ideal candidates for evaluating the utility of voice-based programming support. This motivates the design of our speech-to-code framework, which integrates regional language ASR, LLM-based transcription refinement, and code model response to support accessible, voice-driven code understanding. Moreover, we make our survey details, prompts, and source code publicly available~\cite{souce-code}.

%% file: related_work.tex
\section{Related Work}
Prior work on voice-based programming and programming in native languages remains limited, particularly in the context of code understanding. Most existing systems explore voice as an alternative to text-based code entry, with little attention to downstream reasoning tasks such as code question answering or retrieval.

\textbf{Voice-based code entry.} Early systems such as VoiceGrip~\cite{desilets2001voicegrip} and VoiceCode~\cite{desilets2006voicecode} allow users to dictate code using pseudo-syntax, translating spoken commands into formal programming constructs. These systems improve generic ASR by modeling the unnatural nature of spoken programming syntax, leading to lower symbol error rates. Hossain et al.~\cite{hossain2021code} explore the cognitive load or vocal strain associated with voice-based imperative and reactive programming, observing that while reactive programming reduces vocal effort, it often increases structural complexity. Other systems like~\cite{lagergren2021programming, arnold2000programming} build cross-language abstractions for speech-driven programming, but remain rule-based and English-centric.

\textbf{Voice interaction for software education.} Voice has also been explored for accessible software learning, such as in the Spoken Web project, which enables users to interact with programming environments via conversational commands in regional languages. While promising, these systems focus on high-level composition rather than low-level program comprehension or feedback~\cite{kumar2012spoken}.

\textbf{ASR limitations for code queries.} While general-purpose ASR systems like Whisper~\cite{radford2023robust} have made strides in multilingual transcription, they are not optimized for domain-specific inputs such as code-related speech. Prior studies on ASR robustness in technical domains~\cite{hu2024large} highlight the difficulty of transcribing jargon, identifiers, and punctuation, but do not address downstream consequences for programming tasks.

\textbf{LLM-based ASR refinement.} Post-processing ASR outputs using LLMs is an emerging solution for noisy transcriptions~\cite{adedeji2024sound, ma2025asr}, though most approaches focus on general-purpose sentence correction or captioning. In contrast, our work applies LLMs to refine ASR transcripts specifically for code understanding, where semantic preservation and syntactic accuracy are critical.

\textbf{Multimodal code understanding.} Transformer-based models such as CodeBERT~\cite{feng2020codebert}, CodeT5~\cite{wang2021codet5}, and GPT-4~\cite{openai2024gpt4} have demonstrated strong performance on code summarization, QnA, and retrieval but these models assume clean, text-based queries. Few works explore speech-based access to these capabilities, especially in multilingual and educational settings.

\textbf{Our contribution.} In contrast to prior works, our system targets voice-based \emph{code understanding} rather than code entry. We develop a multilingual framework that accepts spoken queries in regional languages, performs ASR, applies LLM-based \emph{code-aware refinement}, and provides the code model's response. The system delivers text and audio feedback, making it well-suited for novice programmers who may be comfortable with logic but not with English or keyboard-based interaction. To the best of our knowledge, we are the first to work on such an end-to-end framework with promising results.

%% file: framework.tex
\section{Multilingual Speech-to-Code Query Processing Framework}

This section presents an end-to-end framework for processing spoken programming queries in a user’s native language. The proposed system integrates ASR, code-aware LLM refinement, and code generation models to accurately interpret and respond to user queries. Spoken input is first transcribed using ASR, after which the transcription is refined to preserve programming semantics and resolve ambiguities. The refined query is then processed by a code model to generate relevant responses, which are delivered in both textual and spoken formats. This enables accessible and multilingual code assistance.

\subsection{Speech Recognition}

The speech recognition component is responsible for converting spoken user queries into textual form. Given the multilingual nature of the system, an ASR model capable of handling diverse languages and accents is employed. The ASR module produces an initial transcription of the spoken input, which may include programming-related terminology, code keywords, and mixed natural language words. This transcription serves as the foundation for subsequent processing stages, where domain-specific refinement is applied to improve accuracy and preserve code semantics.
Initial experiments using Whisper~\cite{radford2023robust} revealed limitations in handling mixed-language inputs. Specifically, Whisper struggled with queries that have native language utterances with English code terms, which is a common occurrence in student queries. This often led to confusion between phonetically similar tokens, especially when identifiers resembled real words (e.g., \texttt{map}, \texttt{sum}).

\begin{figure}
    \centering
    \includegraphics[width=\linewidth]{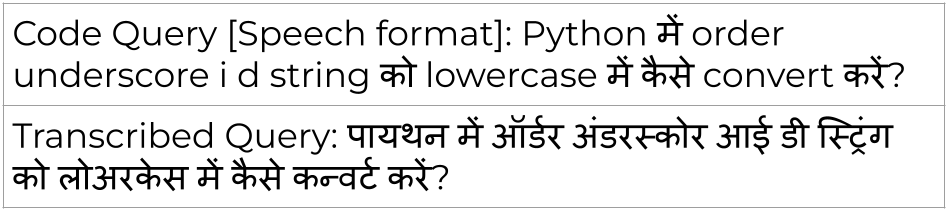}
    \caption{Speech Recognition Example (English Translations for this and other examples are provided in the source code folder)}
    \label{tab:stt_ex}
\end{figure}

To address this, we adopt the \texttt{whisper} for English queries and \texttt{indic-conformer} ASR models for Indic languages' queries. \texttt{indic-conformer} is trained on Indian language speech. These models better preserve mixed-language semantics and improve recognition of technical terms. Table~\ref{tab:stt_ex} shows an example of a spoken Hindi query and its transcription output.

    


\subsection{Code-Aware Transcription Refinement}
\label{subsec:code_aware_refinement}

ASR transcribed code queries may contain errors, particularly for variable names, symbolic operators, and programming keywords. These errors often stem from phonetic ambiguity or misinterpretation of code tokens as natural language. To address this, we introduce a \emph{code-aware transcription refinement} stage that post-processes ASR outputs to improve fidelity and semantic alignment. 

\begin{figure}
    \centering
    \includegraphics[width=\linewidth]{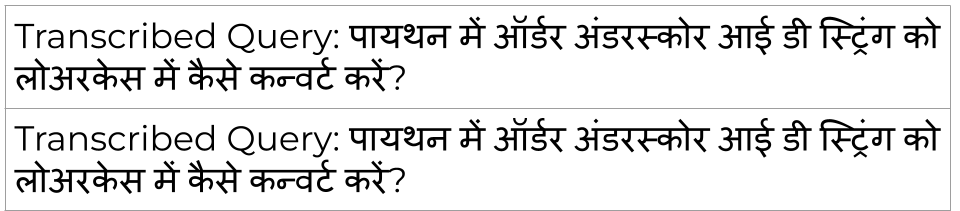}
    \caption{Transcription refinement using code-aware LLM correction}
    \label{tab:error_correction_ex}
\end{figure}

We use a prompt-engineered \textbf{GPT-4o-mini} model to correct transcription errors by leveraging the structure and semantics of programming-related queries. The prompt is designed to: (i) identify and restore misrecognized code terms (e.g., variable names, function calls), (ii) correct phonetically distorted technical words (e.g., ``ask key'' $\rightarrow$ ``ASCII''), (iii) recover symbolic constructs (e.g., \texttt{underscore} $\rightarrow$ \texttt{\_}), and (iv) disambiguate between natural and programming language usage.
This strategy allows the model to reconstruct original code-like phrases even when mispronounced or substituted with similar-sounding non-technical words. It is especially valuable in code-mixed and multilingual scenarios where acoustic overlap is common and exact syntax is critical. Fig~\ref{tab:error_correction_ex} shows an example where the refined transcription accurately restores the intended meaning and structure.

    
    

The refined query, which preserves both the user’s intent and programming semantics, is provided as input to the code model to produce contextually relevant output such as a code snippet, explanation, or debugging suggestion. The model leverages learned programming patterns and domain knowledge to ensure syntactic correctness and semantic coherence. The generated response is then formatted for downstream delivery. This enables both textual presentation and subsequent speech audio in the user’s preferred language.

\subsection{Voice Feedback}
\label{sec:voice_feedback}
In educational and accessibility-focused settings, responses generated by the LLM or code model are converted into spoken form using a Text-to-Speech (TTS) system. This audio delivery closes the voice-based interaction loop, allowing users to submit programming queries through speech and receive spoken responses in their preferred language.

%% file: empirical_setup.tex
\section{Empirical Setup}

We evaluate our voice-based framework for code understanding across multiple programming languages, natural languages, and language models. This section outlines our experimental choices and datasets.

\subsection{Large Language Models}

To explore the impact of model size and capability, we evaluate three LLMs:\\
\textbf{Gemma-9B~\cite{team2024gemma}}: A compact, open-source decoder-only model designed for efficient and lightweight deployment. Gemma demonstrates strong performance on low-resource and Indic natural languages. \\
\textbf{Qwen-30B~\cite{qwen-30b}}: A high-capacity open-weight model with strong reasoning abilities and demonstrated effectiveness on programming and code-related tasks. It is particularly well-suited for code understanding and generation. \\ 
\textbf{GPT-4o-mini~\cite{gpt4omini}}: A frontier commercial model optimized for low-latency inference and multimodal capabilities. As a general-purpose large language model, GPT-4o-mini provides a strong baseline for overall task performance across natural language understanding, multilingual translation, and code comprehension.

    
    

\subsection{Programming and Natural Languages}
We experiment with three widely used programming languages: Python, Java, and PHP, chosen for their diversity and dataset availability. Each code sample is paired with queries in four Indian languages: Hindi, Gujarati, Tamil, and Bengali. These languages span multiple linguistic families and scripts, capturing a broad range of ASR and translation challenges.

\subsection{Datasets}
We use three benchmark datasets:\\
\textbf{(a) CodeQA (QA)~\cite{liu2021codeqa}} A question-answering dataset over source code, designed to test a model’s ability to reason about structure and semantics of code.\\
\textbf{(b) CodeSearchNet (CSN)~\cite{husain1909codesearchnet}} A large-scale code retrieval dataset comprising function-docstring pairs across multiple programming languages. We repurpose it for multilingual code search with natural language queries. \\
\textbf{ (c) CoRNStack (CSk)~\cite{suresh2024cornstack}} A contrastively constructed dataset for code search and reranking. It includes hard negatives and noisy queries, enabling evaluation of robustness in ambiguous query scenarios.

We sample 500 code queries per programming language for each dataset, covering Python, Java, and PHP in CodeSearchNet and CornStack, and Python and Java in CodeQA, resulting in 4,500 unique queries in total. These queries span diverse code constructs, including variables, functions, loops, and conditional statements. Each query is translated into four natural languages, yielding 18,000 multilingual query–code pairs.

\subsection{Multilingual Query Construction}

We use datasets that contain English-language queries paired with source code. Since no multilingual variant exists for Indic languages, we manually construct a query set in four native languages: Hindi, Gujarati, Tamil, and Bengali: through a multi-stage framework: translation, code-aware preprocessing, and text-to-speech synthesis, which is discussed next.

\subsubsection{Translation}

\begin{figure}
    \centering
    \includegraphics[width=\linewidth]{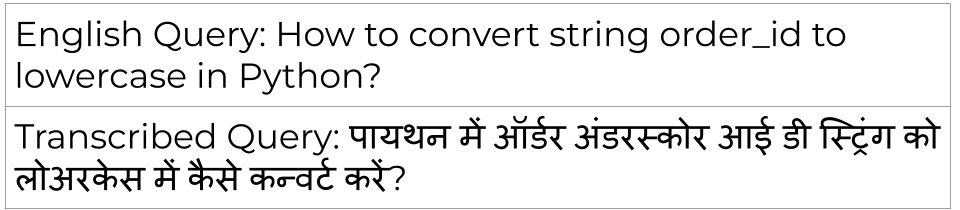}
    \caption{Example: code-aware translation that preserves identifiers}
    \label{tab:translation_example}
\end{figure}

Each English query is translated into one of the four native languages while preserving its semantic meaning and technical structure. As with prior work on task-aware translation~\cite{pan2024lost}, we observe that code-related queries often contain embedded entities such as variable names, function calls, and control structures. These elements are critical to preserve, as mistranslation can distort the functional intent of the query. Generic tools like Google Translate~\cite{googletranslate} are inadequate in this setting; they frequently mistranslate or omit code tokens (e.g., rendering \texttt{print\_sum} as a natural phrase), or fail to distinguish between natural and programming contexts. To address this, we use \textbf{GPT-4o-mini} guided by a carefully designed code-aware prompt. This prompt explicitly instructs the model to: (i) Retain identifiers, variables, and code constructs verbatim (ii) Avoid translating programming keywords and symbolic operators (iii) Preserve grammatical fluency despite mixed-language structure. These constraints ensure the translation remains technically faithful while being readable in the target language. All outputs are manually reviewed to correct inconsistencies and ensure alignment with programming pedagogy. Fig~\ref{tab:translation_example} illustrates an example where code-specific tokens are preserved and embedded naturally in Hindi syntax.



\subsubsection{Code-Aware Preprocessing for TTS}

To ensure compatibility with TTS systems, the translated text is further processed to transform code-related elements into spoken-friendly representations. This involves formatting identifiers, acronyms, and symbols so that they are pronounced correctly by the synthesizer. For instance, \texttt{print\_sum} becomes “print underscore sum” and \texttt{data.user} becomes “data dot user.” Our preprocessing strategy includes:
\begin{itemize}
    \item Spelling out acronyms (e.g., “API” → “A P I”)
    \item Splitting camelCase and snake\_case identifiers \\ (e.g., method name \texttt{getUserInfo} → “get user info”)
    \item Verbalizing special characters using programming conventions (e.g., “==” → “equal equal”)
\end{itemize}

\begin{figure}
    \centering
    \includegraphics[width=\linewidth]{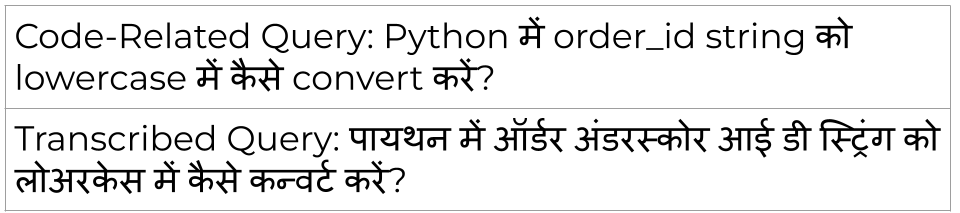}
    \caption{Code-aware formatting for text-to-speech synthesis}
    \label{tab:processed_example}
\end{figure}

These transformations are implemented using a prompt-guided template that ensures user-defined tokens, keywords, and symbolic operators are preserved and clearly rendered in speech (An example is shown in Fig~\ref{tab:processed_example}).


\subsubsection{Text-to-Speech Generation}

The processed text is now passed to a Text-to-Speech (TTS) system to generate audio queries. We evaluate several TTS backends, including GoogleTTS~\cite{googletts}, Gemini 2.5 Preview~\cite{gemini25tts}, and Microsoft Edge TTS~\cite{microsoftedgeTTS}. We select Microsoft Edge TTS due to its strong support for Indian languages, high audio clarity, and open-source accessibility. The TTS system is configured to use native-sounding voices with intelligible pronunciation of both natural language and embedded code fragments.

%% file: llm_effectiveness.tex
\section{LLM Effectiveness in Translation and Transcription Refinement}

LLMs have shown promising capabilities in multilingual text generation and code understanding. In this section, we evaluate their effectiveness in three core stages as part of our framework: (i) code-aware translation from English to Indic languages, (ii) preprocessing for speech synthesis, and (iii) post-ASR transcription refinement. These tasks require precise control over linguistic structure, code semantics, and mixed-language syntax capabilities.

\begin{figure*}[htb]
  \centering
  \includegraphics[width=\textwidth]{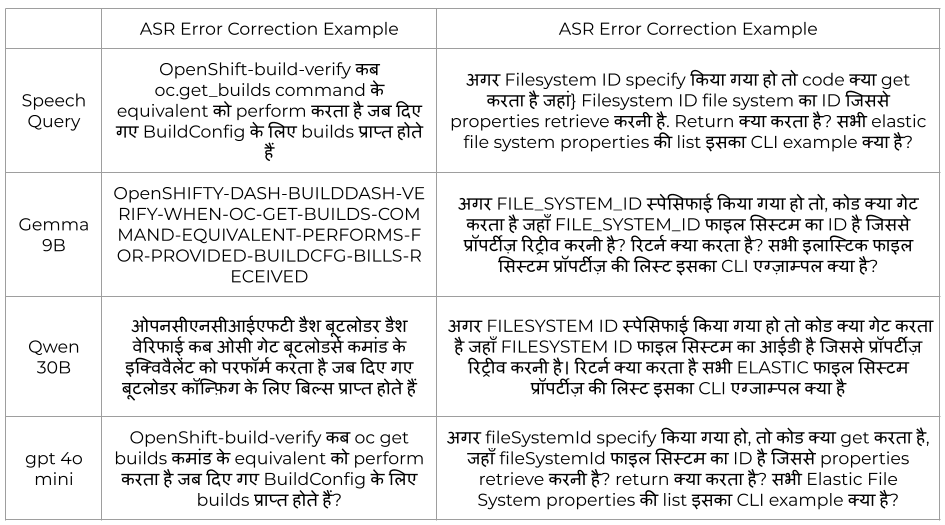} 
  \caption{Refined ASR outputs from different LLMs for speech queries; open-source models struggle with Indic segments.}
  \label{fig:llm_effectiveness}
\end{figure*}

\subsection*{Qualitative Comparison Across LLMs}

Fig~\ref{fig:llm_effectiveness} shows representative examples of LLM outputs for the ASR Error Correction task. We experiment with \textbf{Gemma-9B}, \textbf{Qwen-30B}, and \textbf{GPT-4o-mini}, selected to span a range of model sizes and access constraints. For the translation task, Gemma-9B and Qwen-30B frequently fail to follow code-preserving instructions. They either transliterate English code tokens incorrectly or introduce unnatural literal translations, leading to semantic drift. But GPT-4o-mini consistently produces fluent, code-aware translations that preserve structure and intent.

Similar trends are observed in the preprocessing and error correction stages. Smaller LLMs often ignore symbolic formatting conventions or hallucinate alternate formulations, whereas GPT-4o-mini follows spacing and token rules (e.g., “\texttt{print\_sum}” → “print underscore sum”) more reliably. For ASR refinement, only GPT-4o-mini reliably corrects malformed transcriptions while retaining code semantics and mixed-language phrasing.

\subsection*{Observations and Limitations}

Our empirical findings reveal several failure modes of smaller LLMs in code-related multilingual settings:\\
\textbf{(1) Literal translation}: Direct mappings of English tokens to incorrect native language terms\\
\textbf{(2) code-mixed ambiguity}: Inability to distinguish between natural language and embedded code elements\\
\textbf{(3) Loss of semantic structure}: Degraded or incomplete translations that omit technical meaning

Despite trying multiple open-weight LLMs, we found that only \textbf{GPT-4o-mini} produced usable outputs across all stages in Indic languages. It was the only model to demonstrate consistent semantic preservation, symbolic formatting, and syntactic fluency in code-mixed scenarios. Accordingly, we adopt it as the backbone model for all translation and refinement stages in our framework.

%% file: challenges.tex
\section{Challenges in Multilingual Speech-Based Code Understanding}
\label{sec:challenges}

Speech-based code understanding in multilingual settings introduces complex error modes stemming from accent variation, code-mixed speech, symbolic ambiguity, and syntactic inconsistencies. These errors affect multiple stages in our pipeline—from ASR to translation to text-to-speech. In this section, we qualitatively analyze common failure cases. Later sections provide quantitative evidence of their downstream impact.

\subsection{Speech-to-Text Challenges}

Automatic Speech Recognition (ASR) serves as the first decoding stage, converting spoken multilingual code queries into text. However, it suffers from significant degradation in mixed-language technical scenarios:\\
\textbf{(a) \textit{Phonetic Drift:}}  
English code tokens spoken in regional accents often deviate phonetically. Examples include: (i) “async” $\rightarrow$ “a sink” (ii) “ASCII” $\rightarrow$ “ask key” \\
\textbf{(b) \textit{Code-Mixed Identifier Failures:}} 
CamelCase or snake\_case identifiers like \texttt{getUserInfo} or \texttt{my\_var} are often split, omitted, or substituted entirely. For instance, “getUserInfo” may be transcribed as “get use run foe” or “get your info.” \\
\textbf{(c) \textit{Keyword Ambiguity:}}
Overlapping English words used both in natural and programming language (e.g., “is,” “not,” “in”) are interpreted according to language priors rather than syntactic role. This leads to incorrect transcription of code constructs like “not in” or “is not.” \\
\textbf{(d) \textit{Symbol Loss:}}  
Spoken symbolic operators such as “equal equal” (==) or “not equal” (!=) are rarely preserved. ASR systems either drop them or translate them into incorrect natural phrases (e.g., “barabar barabar”).\\
\textbf{(e) \textit{Identifier Recall Failures:}}  
Identifiers unfamiliar to the ASR model, especially user-defined names or rare abbreviations—are substituted with semantically irrelevant tokens, degrading the downstream reasoning pipeline.\\
\textbf{(f) \textit{Accent based errors:}}
Across the world, people have different accent in pronouncing the words. This may include some english words or even in native language, sometimes people have varying accent. This may lead to mismapping the word to some other word.
examples: (i) “default” (with a strong Marathi accent) $\rightarrow$ “de fall” (ii) “machine” (with a Telugu accent) $\rightarrow$ “mishan”\\
\textbf{(g) \textit{Voice Style Loss:}}
Since ASR systems do not handle voice tones, pauses, emphasis so their importance is completely ignored and hence some errors may happen. 

\subsection{Translation Challenges}

Translation errors compound ASR issues when native language queries must be mapped to semantically aligned, code-aware English text. These challenges are consistent with observations from prior work~\cite{pan2024lost} and further complicated in our setup by the presence of embedded code.\\
\textbf{(a) \textit{Out-of-Vocabulary Entities:}}  
Identifiers in code like \texttt{getData()}, \texttt{APIResponse}, or \texttt{user\_count}  are frequently mistranslated or sometimes transliterated inconsistently. 
\textbf{(b) \textit{Symbol and Syntax Handling:}}  
Programming constructs (e.g., try-catch, ==, !=) are often translated literally. 
\textbf{(c) \textit{Natural vs Code Ambiguity:}} 
Terms like “book,” “is,” “in,” and “override” require contextual disambiguation. 
\textbf{(d) \textit{Grammar and Morphology:}}  
Code-mixed sentences tend to exhibit unnatural grammar in low-resource target languages. Mismatched subject-verb-object order, agreement errors, or partial code token duplication are common.\\
\textbf{(e) \textit{Grammatical Over-correction:}}
Some people may speak queries in informal way or the dataset may contain some informal queries. While translation the model may try to fix the grammar and in that process the query may become semantically slightly different. 
Over-corrected Translation: “It is required to sort the data.”
This Changes intent from a simple task to a necessity.

\subsection{Text-to-Speech Challenges}
Errors in the TTS stage impact pronunciation clarity of both natural and code tokens, reducing the quality of synthesized queries.\\
\textbf{(a) {Identifier Pronunciation:} } 
TTS systems often fail to parse camelCase or acronyms correctly. “checkInk” may be rendered as “checkink,” and “API” may be pronounced as a single word rather than “A P I.”\\
\textbf{(b) \textit{Symbolic Token Misrendering:}} 
Punctuation and programming operators like \texttt{==} and \texttt{!=} are not handled gracefully unless explicitly spaced out or verbalized. This leads to gaps in audio for critical logic operators.\\
\textbf{(c) \textit{Accent and Intonation:}} 
TTS engines trained on standard textual input produce monotonous prosody when reading queries with mixed language and symbols.\\
\textbf{(d) \textit{Lack of Spacing Sensitivity:}}
One of the crucial aspect is code specific words should be handled properly. But while converting to speech, some separate code-specific words or user-defined words may sounds like the joined word. Also, a joint word may sounds like two or more separate words. Since many of such words does not contain in any of natural language's vocabulary so such errors may occur.

%% file: metrics.tex
\section{Evaluation Metrics}

We evaluate our system at two levels: (i) speech recognition accuracy, which reflects how well the ASR component captures spoken input, and (ii) downstream code retrieval performance, which measures how ASR errors affect the final application. Together, these metrics quantify both fidelity and functional impact.

\subsection{Speech Recognition Metrics}

\textbf{Word Error Rate (WER)} is a standard metric for transcription quality:
\[
\text{WER} = \frac{S + D + I}{N}
\]
where $S$, $D$, and $I$ are substitutions, deletions, and insertions at the word level, and $N$ is the number of words in the reference. While informative, WER treats all words equally and fails to capture phonetic similarity, which is particularly problematic when code terms are slightly misheard but acoustically close.

\textbf{Phoneme Error Rate (PER)} addresses limitations of WER by operating on phoneme sequences. We use Epitran~\cite{mortensen2025g2p} to convert transcribed and reference text into phone representations and compute:
\[
\text{PER} = \frac{S_p + D_p + I_p}{N_p}
\]
where $S_p$, $D_p$, and $I_p$ are edit operations at the phoneme level, and $N_p$ is the number of phonemes in the reference. PER is particularly useful for detecting pronunciation-level errors, e.g., when “\texttt{print\_sum}” is transcribed as “\texttt{print sum}” or “\texttt{print some}”, where the acoustic signal is mostly preserved but the code syntax is altered.

\textbf{Weighted Feature-based Edit Distance (WFED)} builds on PER by incorporating articulatory similarity between phonemes. Using PanPhon~\cite{mortensen-etal-2016-panphon}, each IPA symbol is represented as a binary vector of articulatory features (e.g., voicing, nasality, place of articulation). Substitution costs are weighted by phonetic distance, e.g., /s/ $\rightarrow$ /z/ incurs a lower penalty than /s/ $\rightarrow$ /b/. This allows WFED to distinguish minor mispronunciations from more severe acoustic mismatches, offering a finer-grained assessment of how difficult an error is to correct.\\
\textbf{Why PER and WFED Matter:}
In multilingual speech interfaces for code, ASR systems frequently produce outputs that are phonetically close but syntactically incorrect, such as dropping underscores, splitting camelCase tokens, or mispronouncing acronyms. These subtle surface-level errors often render the resulting queries invalid for downstream code models. Unlike WER, which treats all word substitutions equally, PER and WFED provide a more fine-grained view of acoustical faithfulness of ASR output.

Importantly, these metrics help distinguish between recoverable errors, those that a post-ASR correction module can fix, and unrecoverable ones. A low PER or WFED suggests the model recognized the acoustic intent, even if it failed at surface formatting. Conversely, a high PER or WFED indicates that the spoken terms fall outside the model's learned phonetic distribution. In such cases, the ASR system may require domain-specific fine-tuning on code-mixed speech to improve transcription accuracy.

\subsection{Downstream Task Metrics}

To evaluate the functional impact of ASR and correction on real tasks, we use a code retrieval benchmark and report: \textbf{Recall@k} measures whether the correct code snippet is retrieved within the top-$k$ results:
\[
\text{Recall@}k =
\begin{cases}
1, & \text{if ground truth is in top-$k$} \\
0, & \text{otherwise}
\end{cases}
\]

\textbf{Mean Reciprocal Rank (MRR)} captures the position of the correct result:
\[
\text{MRR} = \frac{1}{N} \sum_{i=1}^{N} \frac{1}{\text{rank}_i}
\]

\textbf{Purpose of Downstream Metrics:}
These metrics measure whether transcription quality translates to task utility. A low WER/PER is only meaningful if the query remains semantically useful for retrieval. In our experiments, we use Recall \& MRR to assess whether the ASR output, raw/refined, still allows the code model to retrieve the correct answer. These metrics provide an end-to-end view of how well the speech interface performs.

%% file: results.tex
\section{Result and Analysis}

\begin{table*}[ht]
\centering

\small
\begin{tabular}{p{1.1cm}| p{0.7cm} p{1.1cm}| p{0.6cm} p{0.6cm} p{0.8cm}| p{0.6cm} p{0.6cm} p{0.8cm}| p{0.6cm} p{0.6cm} p{0.8cm}| p{0.6cm} p{0.6cm} p{0.8cm}} 
\textbf{Dataset} & \textbf{Lang} & \textbf{Method} & \multicolumn{3}{c|}{\textbf{Hindi}} & \multicolumn{3}{c|}{\textbf{Gujarati}} & \multicolumn{3}{c|}{\textbf{Tamil}} & \multicolumn{3}{c}{\textbf{Bengali}} \\
 & & & WER & PER & WFED & WER & PER & WFED & WER & PER & WFED & WER & PER & WFED \\
\hline

\textbf{CSN} 
&python & ASR     & 44\% & 33.4\% & 34.5\%  & 43\% & 33.3\%  & 16.7\% & 46\% & 34.1\% & 18.7\% & 64.8\% & 39.7\% & 20.8\%\\
&     &  ASR-R & 30.7\% & 15.4\% & 7.8\%  & 38.6\% & 21.3\% & 11.7\% & 56.6\% & 27.2\% & 17.0\% & 41.9\% & 27.1\% & 14.4\% \\
\cline{2-15}

\multirow{2}{*}
&java & ASR     & 44.7\% & 24.6\% & 19.3\% & 47.6\% & 25.0\% & 20.1\% & 65.6\% & 27.4\% & 19.3\% & 44.0\% & 21.0\% & 14.8\% \\
&  &  ASR-R & 23.8\% & 14.7\% & 7.9\%  & 37.7\% & 28.4\% & 14.2\% & 52.3\% & 25.3\% & 14.9\% & 39.6\% & 22.6\% & 11.9\% \\
\cline{2-15}

\multirow{2}{*}
&php & ASR    & 57.3\% & 40.1\% & 28.0\% & 55.5\% & 42.7\% & 26.7\% & 50.2\% & 32.9\% & 20.9\% & 73.0\% & 42.5\% & 26.9\% \\
&  &  ASR-R & 37.7\% & 28.3\% & 20.0\% & 48.0\% & 32.8\% & 13.7\% & 68.1\% & 36.4\% & 23.2\% & 51.5\% & 30.6\% & 18.3\% \\
\hline

\multirow{2}{*}{\textbf{CSk}} 
&python & ASR      & 46.7\% & 51.0\% & 25.3\% & 44.6\% & 47.2\% & 25.2\% & 47.2\% & 50.6\% & 23.3\% & 54.3\% & 53.5\% & 22.6\% \\
&   &  ASR-R & 31.7\% & 22.9\% & 12.3\% & 36.6\% & 31.4\% & 14.3\% & 52.0\% & 45.1\% & 22.4\% & 42.2\% & 41.7\% & 19.0\% \\
\cline{2-15}
\multirow{2}{*}
&java & ASR     & 48.4\% & 41.9\% & 37.0\% & 56.8\% & 47\% & 30.6\% & 49.5\% & 40.5\% & 34.1\% & 69.0\% & 50.7\% & 42.8\% \\
&  &  ASR-R & 39.0\% & 37.0\% & 26.8\% & 38.9\% & 38.9\% & 14.3\% & 61.8\% & 42.3\% & 36.6\% & 47.8\% & 37.1\% & 27.8\% \\
\cline{2-15}
\multirow{2}{*}
&php & ASR     & 38.8\% & 25.8\% & 14.9\% & 34.5\% & 25.3\%  & 12.6\%  & 39.3\% & 25.0\% & 14.2\% & 61.2\% & 34\% & 20\% \\
& &  ASR-R & 37.7\% & 28.3\% & 20.0\% & 34.5\% & 25.3\% & 12.6\% & 57.4\% & 28.2\% & 17.6\% & 43.1\% & 23.7\% & 12.9\% \\
\hline

\multirow{2}{*}{\textbf{QA}} 
&python & ASR     & 61.4\% & 57.0\% & 34.5\% & 55.8\% & 46.7\% & 18.5\%  & 68.4\% & 45.5\% & 23.3\% & 65.4\% & 44.6\% & 27.6\% \\
&   &  ASR-R & 13.5\% & 3.6\%  & 2.1\%  & 19.4\% & 6.8\%  & 5.3\%  & 43.6\% & 22.3\% & 18.1\% & 49.1\% & 39.0\% & 14.4\% \\
\cline{2-15}
\multirow{2}{*} 
&java & ASR     & 46.7\% & 51.0\% & 25.3\% & 40.8\% & 49.0\% & 25.9\% & 43.8\% & 38.1\% & 21.2\% & 56.6\% & 44\% & 23\% \\
&   &  ASR-R & 24.2\% & 19.6\% & 12.2\% & 31.8\% & 24.8\% & 13.9\% & 45.9\% & 34.3\% & 18.1\% & 39.4\% & 37.7\% & 16.5\% \\
\hline

\end{tabular}
\caption{WER, PER, and WFED (lower is better) across baseline ASR and ASR-R (ASR Refined) queries for the datasets}
\label{tab:main_results}
\end{table*}

We evaluate our multilingual speech-to-code pipeline across three code understanding datasets—CodeSearchNet, CoRNStack, and CodeQA; and four Indic languages: Hindi, Gujarati, Tamil, and Bengali. Our primary evaluation focuses on the ASR task: assessing how well speech inputs are transcribed in the presence of code-specific content. We measure transcription quality using three complementary metrics: Word Error Rate (WER), Phoneme Error Rate (PER), and Weighted Feature-based Edit Distance (WFED). We compare two configurations:
\begin{itemize}
    \item \textbf{ASR:} The baseline output from the speech recognizer
    \item \textbf{Code-Aware Transcription Refinement:} The output after applying our LLM-guided correction step 
\end{itemize}

\subsection{Transcription Accuracy Before and After Refinement}

As shown in Table~\ref{tab:main_results}, across all languages and datasets, raw ASR outputs show high error rates, particularly for Tamil and Bengali, which are two lower-resource languages. WER frequently exceeds 50\%, indicating significant word-level mismatches. PER values often surpass 40\%, reflecting phonetic confusion in recognizing code-related terms. WFED also remains high, often near 25\%.

These elevated error rates underscore the difficulty of transcribing multilingual, code-mixed speech. ASR systems trained on general-purpose corpora struggle to handle identifiers, acronyms, and symbolic operators embedded in natural language. Notably, the high PER values suggest that many spoken forms are acoustically out-of-distribution, indicating that fine-tuning ASR models on code-mixed and technical speech may be necessary to further improve recognition accuracy. 

To address the concern regarding potential model-specific bias, we further evaluated the pipeline by replacing GPT-4o-mini with Claude Sonnet 4.5 for the Error Correction step. This cross-model evaluation allowed us to assess the robustness and generalizability of our approach. Here, we consider 100 samples from the CoRNStack dataset and then follow our pipeline. The results obtained with Claude were as follows (measured in WER, PER, and WFED respectively): Hindi (22.7\%, 7\%, 3.7\%); Gujarati (34.3\%, 10.9\%, 5.3\%); Tamil (48.5\%, 10.8\%, 3.9\%); Bengali (18.2\%, 12.2\%, 5\%). We repeat the same experiment, replacing Claude Sonnet 4.5 with Gemini - 2.5 Pro. The results were: Hindi (20.2\%, 3.8\%, 1.5\%); Gujarati (28.4\%, 15.3\%, 10.9\%); Tamil (49.1\%, 22.5\%, 10.4\%); Bengali (19.3\%, 13.2\%, 5.9\%). These results demonstrate that the pipeline maintains consistent performance trends across different model families, indicating that the observed improvements are not solely dependent on a single model’s characteristics, i.e. a risk optimistic bias.

\subsection{Impact of Code-Aware Transcription Refinement}
Applying our code-aware transcription refinement step using \texttt{GPT-4o-mini} consistently reduces errors across all metrics and languages. On average, we observe a 20.95\% reduction in WER, a 28.85\% reduction in PER, and a 33.36\% reduction in WFED. These improvements demonstrate the effectiveness of our refinement approach in recovering code terms, correcting phonetic distortions, and preserving syntactic structure. Even on individual language breakdowns, the refinement step leads to consistent gains across all metrics.

\subsection{Robustness in Low-Resource Settings}
Notably, these results are achieved in low-resource settings where both speech and NLP resources are limited. The consistent improvements across datasets and languages suggest that code-aware transcription refinement generalizes well and can be deployed to support multilingual programming assistance in underserved regions.

\subsection{Impact on Downstream Code Tasks}
To quantify the downstream effects of ASR errors and the benefit of refinement, we conduct two task-based evaluations using English versions of the datasets, where reference answers are available: \\
\textbf{(a) Question Answering.}  
We prompt a code model to answer queries in three stages: (1) directly using the gold text queries, (2) using ASR-transcribed queries, and (3) using refined queries. To assess answer quality, we use \texttt{GPT-4o-mini} as an evaluator, which compares the predicted answer against the original and categorizes the semantic deviation into three classes: (A) High Deviation, (B) Moderate Deviation, and (C) No Deviation.  
We observe that raw ASR output often introduces errors that significantly alter query intent, leading to wrong or irrelevant answers. Table~\ref{tab:qa_effect} shows a few illustrative cases where code symbols or keywords are misrecognized, resulting in incorrect predictions.\\
\begin{table}[ht]
\centering

\renewcommand{\arraystretch}{1.3}
\small
\begin{tabular}{|p{3.5cm}|p{4.3cm}|}
\hline
\textbf{Original Question} & \textbf{ASR Output} \\
\hline
Why am I getting a double free or corruption error with realloc()? & Why am I getting a double free or corruption error with real lock open close parentheses? \\
\hline
Auto defines in C editors... Why? & Auto defines and see editors.dot.y \\
\hline
Tool to track \#include dependencies & Tool to track cache include dependencies \\
\hline
\end{tabular}
\caption{Comparison of Original and ASR Output Questions}
\label{tab:qa_effect}
\end{table}
\textbf{(b) Code Retrieval}  
We evaluate retrieval performance using the BAAI/bge-code-v1\cite{bdge-code-v1} embedding model. Question embeddings are matched against code embeddings to retrieve top-$k$ results. We compute Recall@$k$ and MRR for three settings: original queries, ASR-transcribed queries, and refined queries. Table~\ref{code_retreival} presents the results. 

Using the original queries, the model achieves 92\% Recall and 87.2\% MRR at $k=5$, and 94\% Recall and 87.51\% MRR at $k=10$. However, with ASR-transcribed queries, performance drops notably to 87\% Recall and 81.58\% MRR at $k=5$, reflecting the disruptive effect of transcription errors. When applying code-aware transcription refinement, the model partially recovers performance, achieving 90\% Recall and 83.78\% MRR at $k=5$, and 92\% Recall and 84.02\% MRR at $k=10$.

\begin{table}[h!]
\centering

\renewcommand{\arraystretch}{1.4}
\begin{tabular}{|>{\raggedright\arraybackslash}m{4cm}|c|c|c|}
\hline
\multirow{2}{*}{\textbf{Method}} & \multicolumn{3}{c|}{\textbf{Metrics}} \\
\cline{2-4}
& \textbf{$k$} & \textbf{Recall}$\uparrow$ & \textbf{MRR}$\uparrow$ \\
\hline
\multirow{2}{*}{Original Text Query} 
& 5  & 92\%  & 87.2\% \\
& 10 & 94\%  & 87.51\% \\
\hline
\multirow{2}{*}{ASR Transcribed Query} 
& 5  & 87\%  & 81.58\% \\
& 10 & 88\%  & 81.7\% \\
\hline
\multirow{2}{*}{Code-Aware Refined Query} 
& 5  & 90\%  & 83.78\% \\
& 10 & 92\%  & 84.02\% \\
\hline
\end{tabular}
\caption{Recall and MRR at $k$ for BAAI/bge-code-v1}
\label{code_retreival}
\end{table}

\subsection{Summary}
\label{summary}
These results confirm that transcription errors have a measurable and often detrimental impact on downstream code tasks, even when those errors are phonetically minor (\textbf{RQ1}). We observe that ASR systems frequently introduce structural distortions such as dropped underscores, mistranscribed symbols, and substitution of code terms with phonetically similar non-technical words. These patterns represent recurring failure modes in code-mixed transcription. These challenges are discussed in detail in Section~\ref{sec:challenges} (\textbf{RQ2}).

Performance degradation is pronounced in low-resource Indic languages like Tamil and Gujarati, where higher WER, PER, and WFED scores indicate both acoustic and syntactic mismatch (\textbf{RQ3}). While general-purpose LLMs like GPT-4o-mini show strong capabilities in text-based reasoning, they are not inherently robust to ASR-induced noise. Our proposed code-aware transcription refinement module significantly improves both transcription quality and the performance of downstream code understanding tasks. This clearly demonstrates the effectiveness of LLM-guided error correction in recovering the semantic and syntactic integrity of spoken queries (\textbf{RQ4}).

These findings reinforce the need for ASR post-processing in speech-based programming systems, particularly in educational contexts targeting multilingual and low-resource user populations.

%% file: multimodal.tex
\section{Comparison with Multimodal LLMs}
\label{phi-4}

\begin{table}[h!]
\centering

\begin{tabular}{|c|ccc|}
\hline
\multirow{2}{*}{\textbf{Method}} & \multicolumn{3}{c|}{\textbf{Metrics}} \\
\cline{2-4}
 & \textbf{WER}$\downarrow$ & \textbf{PER}$\downarrow$ & \textbf{WFED}$\downarrow$ \\
\hline
\textbf{Phi-4} & 27.9\% & 9.3\% & 7.4\% \\
\textbf{Qwen3-Omni-Flash} & 39.6\% & 8.1\% & 7.3\% \\
\textbf{Our Framework} & 19.9\% & 2.9\% & 1.3\% \\
\hline
\end{tabular}
\caption{Comparison of Phi-4 and our framework}
\label{tab:phi4}
\end{table}

To contextualize the effectiveness of our ASR cascade pipeline, we compare it against a recent end-to-end multimodal large language model: \textbf{Phi-4-multimodal-instruct}~\cite{phi-4}. Unlike our approach, which explicitly separates speech recognition and LLM-based refinement, Phi-4 directly consumes raw audio inputs and generates transcriptions in a single, non-cascaded pass. This design eliminates intermediate ASR stages but also forgoes the benefits of domain-specific correction. We evaluate on a manually curated set of code-related spoken queries. As shown in Table~\ref{tab:phi4}, Phi-4 achieves a WER of 27.9\%, PER of 9.3\%, and WFED of 7.4\%. Similarly, Qwen-3-Omni-Flash  achieves a WER of 39.6\%, PER of 8.1\%, and WFED of 7.3\%. In contrast, our code-aware ASR cascade framework substantially improves performance, reducing WER to 19.9\%, PER to 2.9\%, and WFED to 1.3\%. 

These results suggest that even state-of-the-art multimodal models struggle with transcribing technical, code-mixed utterances. Moreover, Phi-4 does not currently support Indic languages, and the above evaluation is conducted solely on English inputs. This highlights a critical limitation of existing end-to-end multimodal LLMs: they are not yet robust to multilingual, code-intensive speech, nor do they provide the modular flexibility required for correction or adaptation. In contrast, our framework explicitly addresses these challenges and achieves significantly lower error rates across all transcription metrics.

Furthermore, the framework was deployed within College X’s LMS platform, to which we had authorized access, and conducted a demonstration trial session in a computer lab. This helped us evaluate the effectiveness and practical usefulness of our framework. A total of 28 participants, primarily beginners in coding, were invited to interact with our framework, which was deployed. They were asked to rate their overall experience with our system. The responses were: Excellent (10.7\%), Good (32.1\%), Satisfactory (28.6\%), Fair (10.7\%), and Needs Improvement (17.9\%). These results indicate that more than 80\% of participants rated their experience as satisfactory or above, highlighting the potential of our system as a supportive tool for beginners. In addition to rating their experience, participants were asked (26 participants were involved): \textit{``If your concerns are resolved, would you like to use our system?''} and options: \textit{A. Yes, definitely, B. Yes, probably, C. Maybe, D. Probably not, E. Definitely not}. Responses were as follows: 14 selected A, 11 selected B, 1 selected C, 0 selected D \& E. The survey results demonstrate that the framework is effective not only for TTS-generated audio but also for real-time voice queries in noisy environments.

%% file: validity.tex
\section{Threats to Validity}

We now discuss key threats to the validity of our results.

\subsection{External Validity}


External validity concerns the generalizability of our findings across languages, datasets, and models. To address this, we used three common programming languages (Python, Java, PHP) and diverse benchmarks: CodeQA, CodeSearchNet, and CoRNStack along with two strong LLMs, GPT-4o-mini, Phi-4-multimodal-instruct, and Qwen3-Omni-Flash ensuring broad applicability.

\subsection{Internal Validity}
A potential threat to internal validity arises from differences between controlled experimental conditions and real-world usage scenarios. Although the framework was evaluated through a live demonstration involving real users and voice-based interactions, certain components of the system, such as synthesized speech outputs and controlled laboratory settings, may not fully capture the diversity of natural speaking styles, accents, pacing, and spontaneous phrasing encountered in broader deployments. While these choices support consistency during evaluation, further studies in less controlled environments and with more diverse user populations are needed to confirm the robustness of the framework under real-world conditions.



%% file: conclusion.tex
\section{Conclusion and Future Work}

We present a modular, speech-driven code understanding framework for multilingual learners, focusing on Indic languages. The system integrates ASR, code-aware transcription refinement, and LLM-based reasoning to help students learn programming in their native languages. Evaluations on CodeSearchNet, CoRNStack, and CodeQA reveal that ASR errors, especially in low-resource, code-mixed settings, significantly hinder performance, while LLM-guided refinement greatly enhances transcription quality and downstream accuracy.

In future, we plan to fine-tune ASR models on code-mixed, domain-specific speech to reduce phoneme errors, extend coverage to more Indian languages through scalable augmentation and integrate with real-time educational tools to enable multilingual, voice-based programming tutors for classroom and mobile use, further advancing inclusive and accessible programming education. Finally, while the current system focuses on Indian regional languages, we aim to extend support to widely used international languages, including French, Arabic, Spanish, and German, to further broaden accessibility.

%% file: reference.bib
@misc{cursor_ai,
  author = {{Cursor}},
  title = {{Cursor: The AI Code Editor}},
  howpublished = {\url{https://cursor.sh/}},
  note = {Accessed: 2025-09-29}
}

@misc{visual_studio_code,
  author = {{Microsoft}},
  title = {{Visual Studio Code}},
  howpublished = {\url{https://code.visualstudio.com/}},
  note = {Accessed: 2025-09-29}
}

@inproceedings{desilets2006voicecode,
  title={Voicecode: An innovative speech interface for programming-by-voice},
  author={D{\'e}silets, Alain and Fox, David C and Norton, Stuart},
  booktitle={CHI'06 extended abstracts on Human factors in computing systems},
  pages={239--242},
  year={2006}
}

@article{desilets2001voicegrip,
  title={VoiceGrip: a tool for programming-by-voice},
  author={Desilets, Alain},
  journal={International Journal of Speech Technology},
  volume={4},
  number={2},
  pages={103--116},
  year={2001},
  publisher={Springer}
}

@inproceedings{hossain2021code,
  title={Code generator based on voice command for multiple programming language},
  author={Hossain, Sakib and Emi, Mabia Akter and Mishu, Mohsina Hossain and Zannat, Raihana and others},
  booktitle={2021 12th International Conference on Computing Communication and Networking Technologies (ICCCNT)},
  pages={01--05},
  year={2021},
  organization={IEEE}
}

@misc{lagergren2021programming,
  title={Programming by voice: Efficiency in the Reactive and Imperative Paradigm},
  author={Lagergren, Miriam and Soneryd, Max},
  year={2021}
}

@inproceedings{arnold2000programming,
  title={Programming by voice, VocalProgramming},
  author={Arnold, Stephen C and Mark, Leo and Goldthwaite, John},
  booktitle={Proceedings of the fourth international ACM conference on Assistive technologies},
  pages={149--155},
  year={2000}
}

@misc{googletranslate,
  author = {{Google}},
  title = {{Google Translate}},
  howpublished = {\url{https://translate.google.com/}},
  note = {Accessed: 2025-07-16}
}

@misc{googletts,
  author = {{Google Cloud}},
  title = {{Google Cloud Text-to-Speech}},
  howpublished = {\url{https://cloud.google.com/text-to-speech}},
  note = {Accessed: 2025-07-16}
}

@misc{gemini25tts,
  author = {{Google DeepMind}},
  title = {{Gemini 2.5 Preview – Text-to-Speech (TTS)}},
  howpublished = {\url{https://ai.google.dev/gemini-api/docs/models\#gemini-2.5-pro-preview-tts}},
  note = {Accessed: 2025-07-16}
}

@misc{microsoftedgeTTS,
  author = {{Microsoft Corporation}},
  title = {{Microsoft Edge Text-to-Speech (TTS)}},
  howpublished = {\url{https://microsoftedge.microsoft.com/addons/detail/read-aloud-a-text-to-spe/pnfonnnmfjnpfgagnklfaccicnnjcdkm}},
  note = {Accessed: 2025-07-16}
}

@misc{qwen-30b,
  author = {{Qwen Team., Alibaba}},
  title = {{Qwen3-30B-A3B}},
  howpublished = {\url{https://huggingface.co/Qwen/Qwen3-30B-A3B}},
  note = {Accessed: 2025-07-16}
}

@misc{gpt4omini,
  author = {{OpenAI}},
  title = {{gpt-4o-mini}},
  howpublished = {\url{https://openai.com/index/gpt-4o-mini-advancing-cost-efficient-intelligence/}},
  note = {Accessed: 2025-07-16}
}

@misc{bdge-code-v1,
  author = {{BAAI}},
  title = {{BAAI/bge-code-v1}},
  howpublished = {\url{https://huggingface.co/BAAI/bge-code-v1}},
  note = {Accessed: 2025-07-16}
}

@misc{souce-code,
  author = {{Anonymous}},
  title = {{Source Code}},
  howpublished = {\url{https://drive.google.com/drive/folders/1x_PwXTyf52ZwtC1vWdO5kQmsGacunNCv}},
  note = {Accessed: 2025-07-18}
}

@article{phi-4,
  title={Phi-4-mini technical report: Compact yet powerful multimodal language models via mixture-of-loras},
  author={Abouelenin, Abdelrahman and Ashfaq, Atabak and Atkinson, Adam and Awadalla, Hany and Bach, Nguyen and Bao, Jianmin and Benhaim, Alon and Cai, Martin and Chaudhary, Vishrav and Chen, Congcong and others},
  journal={arXiv preprint arXiv:2503.01743},
  year={2025}
}

@inproceedings{reitmaier2022opportunities,
  title={Opportunities and challenges of automatic speech recognition systems for low-resource language speakers},
  author={Reitmaier, Thomas and Wallington, Electra and Kalarikalayil Raju, Dani and Klejch, Ondrej and Pearson, Jennifer and Jones, Matt and Bell, Peter and Robinson, Simon},
  booktitle={Proceedings of the 2022 CHI conference on human factors in computing systems},
  pages={1--17},
  year={2022}
}

@inproceedings{nigatu2024low,
  title={Low-Resourced Languages and Online Knowledge Repositories: A Need-Finding Study.},
  author={Nigatu, Hellina Hailu and Canny, John and Chasins, Sarah E},
  booktitle={Proceedings of the 2024 CHI Conference on Human Factors in Computing Systems},
  pages={1--21},
  year={2024}
}

@inproceedings{paudyal2020voiceye,
  title={Voiceye: A multimodal inclusive development environment},
  author={Paudyal, Bharat and Creed, Chris and Frutos-Pascual, Maite and Williams, Ian},
  booktitle={Proceedings of the 2020 ACM Designing Interactive Systems Conference},
  pages={21--33},
  year={2020}
}

@inproceedings{paudyal2022inclusive,
  title={Inclusive multimodal voice interaction for code navigation},
  author={Paudyal, Bharat and Creed, Chris and Williams, Ian and Frutos-Pascual, Maite},
  booktitle={Proceedings of the 2022 International Conference on Multimodal Interaction},
  pages={509--519},
  year={2022}
}

@inproceedings{unni2022adaptive,
  title={Adaptive discounting of implicit language models in rnn-transducers},
  author={Unni, Vinit and Khare, Shreya and Mittal, Ashish and Jyothi, Preethi and Sarawagi, Sunita and Bharadwaj, Samarth},
  booktitle={ICASSP 2022-2022 IEEE International Conference on Acoustics, Speech and Signal Processing (ICASSP)},
  pages={8122--8126},
  year={2022},
  organization={IEEE}
}

@inproceedings{unni2023improving,
  title={Improving RNN-Transducers with Acoustic LookAhead},
  author={Unni, Vinit S and Mittal, Ashish and Jyothi, Preethi and Sarawagi, Sunita},
  booktitle={Proc. Interspeech 2023},
  pages={4419--4423},
  year={2023}
}

@inproceedings{radford2023robust,
  title={Robust speech recognition via large-scale weak supervision},
  author={Radford, Alec and Kim, Jong Wook and Xu, Tao and Brockman, Greg and McLeavey, Christine and Sutskever, Ilya},
  booktitle={International conference on machine learning},
  pages={28492--28518},
  year={2023},
  organization={PMLR}
}

@inproceedings{pan2024lost,
  title={Lost in translation: A study of bugs introduced by large language models while translating code},
  author={Pan, Rangeet and Ibrahimzada, Ali Reza and Krishna, Rahul and Sankar, Divya and Wassi, Lambert Pouguem and Merler, Michele and Sobolev, Boris and Pavuluri, Raju and Sinha, Saurabh and Jabbarvand, Reyhaneh},
  booktitle={Proceedings of the IEEE/ACM 46th International Conference on Software Engineering},
  pages={1--13},
  year={2024}
}

@article{kumar2012spoken,
  title={Spoken web: using voice as an accessibility tool for disadvantaged people in developing regions},
  author={Kumar, Arun and Agarwal, Sheetal K},
  journal={ACM SIGACCESS Accessibility and Computing},
  number={104},
  pages={3--11},
  year={2012},
  publisher={ACM New York, NY, USA}
}

@article{hu2024large,
  title={Large language models are efficient learners of noise-robust speech recognition},
  author={Hu, Yuchen and Chen, Chen and Yang, Chao-Han Huck and Li, Ruizhe and Zhang, Chao and Chen, Pin-Yu and Chng, EnSiong},
  journal={arXiv preprint arXiv:2401.10446},
  year={2024}
}

@article{ma2025asr,
  title={Asr error correction using large language models},
  author={Ma, Rao and Qian, Mengjie and Gales, Mark and Knill, Kate},
  journal={IEEE Transactions on Audio, Speech and Language Processing},
  year={2025},
  publisher={IEEE}
}

@article{adedeji2024sound,
  title={The sound of healthcare: Improving medical transcription asr accuracy with large language models},
  author={Adedeji, Ayo and Joshi, Sarita and Doohan, Brendan},
  journal={arXiv preprint arXiv:2402.07658},
  year={2024}
}

@misc{feng2020codebert,
      title={CodeBERT: A Pre-Trained Model for Programming and Natural Languages}, 
      author={Zhangyin Feng and Daya Guo and Duyu Tang and Nan Duan and Xiaocheng Feng and Ming Gong and Linjun Shou and Bing Qin and Ting Liu and Daxin Jiang and Ming Zhou},
      year={2020},
      eprint={2002.08155},
      archivePrefix={arXiv},
      primaryClass={cs.CL},
      url={https://arxiv.org/abs/2002.08155}, 
}

@misc{wang2021codet5,
      title={CodeT5: Identifier-aware Unified Pre-trained Encoder-Decoder Models for Code Understanding and Generation}, 
      author={Yue Wang and Weishi Wang and Shafiq Joty and Steven C. H. Hoi},
      year={2021},
      eprint={2109.00859},
      archivePrefix={arXiv},
      primaryClass={cs.CL},
      url={https://arxiv.org/abs/2109.00859}, 
}

@misc{openai2024gpt4,
      title={GPT-4 Technical Report}, 
      author={OpenAI and Josh Achiam and Steven Adler and Sandhini Agarwal and Others},
      year={2024},
      eprint={2303.08774},
      archivePrefix={arXiv},
      primaryClass={cs.CL},
      url={https://arxiv.org/abs/2303.08774}, 
}

@article{team2024gemma,
  title={Gemma 2: Improving open language models at a practical size, 2024},
  author={Team, Gemma and Riviere, Morgane and Pathak, Shreya and Sessa, Pier Giuseppe and Hardin, Cassidy and Bhupatiraju, Surya and Hussenot, L{\'e}onard and Mesnard, Thomas and Shahriari, Bobak and Ram{\'e}, Alexandre and others},
  journal={URL https://arxiv. org/abs/2408.00118},
  volume={1},
  number={3},
  year={2024}
}

@article{liu2021codeqa,
  title={CodeQA: A question answering dataset for source code comprehension},
  author={Liu, Chenxiao and Wan, Xiaojun},
  journal={arXiv preprint arXiv:2109.08365},
  year={2021}
}

@article{husain1909codesearchnet,
  title={CodeSearchNet Challenge: Evaluating the State of Semantic Code Search.(2019)},
  author={Husain, Hamel and Wu, Ho-Hsiang and Gazit, Tiferet and Allamanis, Miltiadis and Brockschmidt, Marc},
  journal={arXiv preprint arXiv:1909.09436},
  year={1909},
  publisher={CoRR}
}

@article{suresh2024cornstack,
  title={CoRNStack: High-quality contrastive data for better code retrieval and reranking},
  author={Suresh, Tarun and Reddy, Revanth Gangi and Xu, Yifei and Nussbaum, Zach and Mulyar, Andriy and Duderstadt, Brandon and Ji, Heng},
  journal={arXiv preprint arXiv:2412.01007},
  year={2024}
}

@article{mortensen2025g2p,
  title={G2P and P2G},
  author={Mortensen, David R},
  year={2025}
}

@inproceedings{mortensen-etal-2016-panphon,
    title = "{P}an{P}hon: A Resource for Mapping {IPA} Segments to Articulatory Feature Vectors",
    author = "Mortensen, David R.  and
      Littell, Patrick  and
      Bharadwaj, Akash  and
      Goyal, Kartik  and
      Dyer, Chris  and
      Levin, Lori",
    editor = "Matsumoto, Yuji  and
      Prasad, Rashmi",
    booktitle = "Proceedings of {COLING} 2016, the 26th International Conference on Computational Linguistics: Technical Papers",
    month = dec,
    year = "2016",
    address = "Osaka, Japan",
    publisher = "The COLING 2016 Organizing Committee",
    url = "https://aclanthology.org/C16-1328/",
    pages = "3475--3484",
}
